\documentstyle[HEP93]{article}
\begin{document}
\newcommand{\newc}{\newcommand}
\newc{\gev}{\,GeV}
\newc{\ra}{\rightarrow}
\newc{\lsim}{\buildrel{<}\over{\sim}}
\newc{\gsim}{\buildrel{>}\over{\sim}}
\newc{\esim}{\buildrel{\sim}\over{-}}
\newc{\lam}{\lambda}
\newc{\mpl}{M_{Pl}}
\newc{\mw}{M_W}
\newc{\pho}{{\tilde\gamma}}
\newc{\half}{\frac{1}{2}}
\newc{\third}{\frac{1}{3}}
\newc{\quarter}{\frac{1}{4}}
\newc{\beq}{\begin{equation}}
\newc{\eeq}{\end{equation}}
\newc{\barr}{\begin{eqnarray}}
\newc{\earr}{\end{eqnarray}}
\newc{\cl}{\centerline}
\newc{\delb}{\Delta B\not=0}
\newc{\dell}{\Delta L\not=0}
\newc{\delbi}{\Delta B_i\not=0}
\newc{\delli}{\Delta L_i\not=0}
\newc{\lra}{\leftrightarrow}
\title{ Lepton Masses protect the Primordial Baryogenesis from Sphaleron
Erasure}
\author{H.~Dreiner \\
Theoretical Physics, ETH Z\"urich, CH-8093 Z\"urich}
\abstract{ We present a
revision of the analysis of sphaleron baryon-number violating processes in the
standard model including lepton-mass effects. We find the surprising result
that
a GUT-scale matter-asymmetry can survive the $B$ and $L$ violating sphaleron
interactions even though ($B- L$) is conserved and equals zero for all
temperatures. We extend the analysis to cover the minimal supersymmetric
standard model (MSSM). As a new result we present the MSSM analysis below
the temperature of electroweak breaking and for $B-L=0$ at all
temperatures$^1$.}
\maketitle

It has been observed  that the
requirement of a cosmological baryon-asymmetry leads to powerful constraints on
the initial nature of the baryon asymmetry.
The basic point is that Sphaleron
induced processes are thought to be in thermal equilibrium above the
electroweak breaking scale leading to the possible erasure of any pre-existing
baryon-number excess. They  conserve $(B-L)$ and so, it is normally
argued, any baryon-number excess produced at an early epoch by a $(B-L)$
conserving interaction  will be erased. Thus $(B-L)$ violating processes at the
GUT scale are apparently needed if the baryon-number asymmetry is to survive.
Here we re-examine these ideas.\footnote{This is a very concise
summary of work performed in collaboration
with G.G. Ross Ref.\cite{paper}. The only new result is given in
Eq.(\ref{eq:new}). Throughout we shall use the notation of
\cite{paper}.} Going beyond previous analyses \cite{khs}, we include all mass
corrections including the lepton-masses. Surprisingly,
these  lead to very significant effects; for example
a GUT-scale matter asymmetry created in a B=L channel {\it can } survive the
sphaleron interactions. We extend this idea to
supersymmetry and show that slepton mass effects are much larger and offer
a more promising possibility to protect a GUT-scale $B=L$ mattergenesis.

The gauge interactions are in thermal equilibrium well below $T=\mw$.
The Sphaleron  \cite{ht6} interactions violate baryon-number $B_i$
and lepton-number $L_i$ but conserve $\third B-L_i,\,
B_i-B_j,$ of which five are linearly independent. They are in thermal
equilibrium for all temperatures\footnote{$T_{sphal}$ is the lowest
temperature where the sphalerons are still in equilibrium.} $T> T_{sphal}$.

The off-diagonal quark interactions violate all
$B_i$, but preserve the total baryon-number $B$ and are all in thermal
equilibrium
for $T> T_{sphal}$ \cite{paper}. Thus, in the SM the
symmetries are reduced to the three conserved quantum numbers
\begin{equation}
\third B-L_i.
\label{eq:cons3}
\end{equation}
Given the separate conserved lepton numbers it is clear from
Eq.(\ref{eq:cons3}) that a pre-existing baryon number survives provided there
is a component produced in a $(\third B- L_{i})=0$ channel. It should also
be clear that $B-L=0$ at all temperatures does not necessarily imply
$(\third B- L_{i})=0$, and thus a baryonasymmetry can in principle survive.
We shall see in the chemical potential analysis how this happens.

The net number density $n_k$ of a given particle
species is approximately given by \cite{weinbook}
\begin{eqnarray}
n_k & \approx & \frac{g_k}
{\pi^2} T^3 \left(\frac{\mu_k}{T}\right) \int_{m_k/T}^\infty
    y \sqrt{y^2- m_k^2/T^2} \frac{e^y}{(1\pm e^y)^2} dy \nonumber \\
&\equiv& \frac{g_k} {\pi^2} T^3 \left( \frac {\mu_k}{T}
\right) F_\pm( \frac          {m_k}{T}),
\label{eq:chempot}
\end{eqnarray}
proportional to $\frac{g_k} {\pi^2} T^3 ({\mu_k}/{T})$. We have assumed $\mu/
T\ll 1$. We denote the mass dependent proportionality factor
\begin{equation}
{F_\pm(\frac{m_k}{T})} = {F_\pm(0)} \alpha(\frac{m_k}{T}).
\end{equation}
The introduction of $\alpha\not=1$ for non-zero masses is the deceisive
new point in our analysis.

At {\it sufficiently high temperatures and densities}
the chemical potentials of the photon, gluon and $Z^0$ vanish.
Therefore the chemical potentials of particles and antiparticles are equal and
opposite and different coloured quarks have equal chemical
potentials. The off-diagonal Yukawa interactions
guarantee that the chemical potential
of all up-like quarks and all down-like quarks are respectively equal. We are
thus left with $3N+7$ chemical potentials. It is convenient to
introduce the following notation ($\mu_i=\mu_{\nu_i}$)
\begin{equation}
\begin{array}{rcl}
\alpha_i &\equiv&  \alpha({m_{e_i}}/ {T}),\\
 \Delta_i &\equiv&N-\sum_i \alpha_i,  \\
  \mu &\equiv& \sum_i\mu_i,\\
 {\bar\mu} &=&\sum_i\alpha_i\mu_i, \\
\Delta\mu &=& \mu-{\bar\mu}, \\
\Delta_u &\equiv& N- \sum_i \alpha( {m_{u_i}}/ {T_{sphal}}), \\
\Delta_d &\equiv & N- \sum_i \alpha( {m_{d_i}}/ {T_{sphal}}), \\
\alpha_- &\equiv&  \alpha( {m_{\phi^-}} /{T}), \\
\alpha_0 &\equiv& \alpha({m_{\phi^0}} /{T}), \\
 \alpha_W &\equiv& \alpha({T_{sphal}}/{T}).
\end{array}
\label{eq:chemdef}
\end{equation}
In the massless limit $\Delta\mu=\Delta_i= \Delta_u= \Delta_d=0$, $\alpha(0)=1
$. At $T_{sphal}=\mw$ and for $m_{top}=150\gev$: $\Delta_i,\Delta_d\ll
\Delta_u \approx 0.38$. We thus neglect both $\Delta_i$ and $\Delta_d$.
The electroweak interactions lead to the following $4+2N=10$ equilibrium
relations among the chemical potentials
\begin{equation}
\begin{array}{llllll}
\mu_W &=& \mu_-+ \mu_0  &\mu_{dL} &=& \mu_{uL}+ \mu_W
\\
\mu_{iL} &=& \mu_i+ \mu_W    &
\mu_{uR} &=& \mu_0+ \mu_{uL}
\\  \mu_{dR} &=& -\mu_0+ \mu_W + \mu_{uL}  &
  \mu_{iR} &=& -\mu_0+ \mu_W+\mu_i
\end{array}
\label{eq:chemeq}
\end{equation}
independent of the mass corrections $\alpha$ and internal degrees of freedom $
g_k$ \cite{weinbook}. However, the net value of a quantum number depends on the
net number density and thus depends on the product of $\mu$, $g_i$ and most
importantly $\alpha$.
Using the equations (\ref{eq:chemeq}), we can express all of the chemical
potentials in terms of $6$, which we chose to be $\mu_W,\mu_0,\mu_{uL},$ and $
\mu_i$. However, $\mu_i$ only appears in the combinations $\mu$ and $\Delta\mu
$ leaving only 5 independent chemical potentials.

We can now express the total electric charge $Q$ and the third component of
weak isospin $Q_3$, as well as $B,L$ in terms of these 5 chemical potentials,
{\it e.g.}
\begin{eqnarray}
Q &=&  2(N-2\Delta_u) \mu_{uL}  - 2(2N+ 2\alpha_W +b\alpha_-)
\mu_W \nonumber \\
&&   -2(\mu-\Delta\mu)   +2(2N- \Delta_u + b\alpha_-)\mu_0
\label{eq:charge} \\
Q_3&=&-\frac{3}{2}\Delta_u\mu_{uL}-(2N+4\alpha_W+b\alpha_-)\mu_W
\nonumber \\
&&+ b(\alpha_- -\alpha_0)\mu_0 +\frac{1}{2}\Delta\mu , \\
B&=& (4N-2\Delta_u)\mu_{uL}+2N\mu_W-\Delta_u\mu_0,\\
L&=& 3\mu-2\Delta\mu+2N\mu_W-N\mu_0.
\end{eqnarray}
We have for $T> T_{sphal}$ one further equation due to the Sphaleron
interactions
\beq
N(3\mu_{uL}+2\mu_W)+\mu =0.
\label{eq:chemsphal}
\eeq
At all temperatures $U(1)_Q$ is a good symmetry and therefore we must have $Q=
const.\approx 0$ \cite{lyttle}. Above $T_C$, we also have $Q_3=0$. For $T<T_C$,
$Q_3\not=0$ but $<\phi>\not=0$ which implies
$\mu_0=0$. Thus for $T> T_{sphal} $, we have 3 equations beyond those of
Eq.(\ref{eq:chemeq}), for the five unknowns:
$\mu_{uL},\,\mu_W,\,\mu_0,\mu,$ and $ \Delta\mu$. Hence, we can write $B$ and
$L$ in terms of $\mu_{uL}$ and $\Delta\mu$.

(1){$\bf T\gsim T_C$}
Above the scale for electroweak breaking the quarks, leptons and the
W-boson are massless, $\Delta_u= \Delta_d= \Delta_i=0$, $\alpha_W=\alpha_i=
1$, and $\Delta\mu=0$. We can thus write $B$ and $L$ in terms
of $\mu_{uL}$ only
\beq
B=4N\mu_{uL}, \quad L=- \frac {14N^2+9Nb\alpha_-} {2N+b\alpha_-} \mu_{uL}.
\label{eq:baryhigh}
\eeq
Above $T_C$ the mass corrections do not modify the value of $B$ and only
very mildly modify $L$. From Eq.(\ref{eq:baryhigh}) and the observational
value for
$n_B/s$  we obtain $\mu_{uL}\approx 10^{-11}$. Also
\begin{eqnarray}
B+L&=& -\frac {6N^2+ 5Nb\alpha_-} {2N+b \alpha_-} \mu_{uL} \nonumber \\
&=& -\frac
{6N+5b\alpha_-} {22N+13b\alpha_-} (B-L),
\end{eqnarray}
Thus a non-zero value for $B-L$ implies a non-zero value
for $B+L$, even though the $B+L$ violating Sphaleron interactions are in
thermal equilibrium.

(2){ $\bf T_{sphal} \lsim T \lsim T_C$}
Setting $\mu_0=0$ and using Eqs.(\ref{eq:charge}) we obtain
\begin{eqnarray}
B&=& \left(4N- 2\Delta_u +\frac {4N (2N- \Delta_u) } {2\alpha_W
+b\alpha_-} \right)     \mu_{uL}  \\
&& + \frac{2N} {2\alpha_W+ b
\alpha_-} \Delta\mu  , \\  %
L&=& -\left( 9N+ \frac {8N (2N- \Delta_u )} {2\alpha_W+b \alpha_-
}  \right)      \mu_{uL} \\
&&- 2\left( 1+ \frac {2N} {2\alpha_W+
b\alpha_-}  \right) \Delta\mu,
\label{eq:now}
\end{eqnarray}
In the massless limit Eqs.(\ref{eq:baryhigh}-\ref{eq:now}) agree with
\cite{harvey}.
We find a non-zero value for $B+L$, which is not washed out by the sphalerons.
The mass corrections due to the top quark are
about $5\%$, the correction due to the Higgs and $W$-mass is about a factor of
two. The more important mass effect is that $B+L$ is {\it not} proportional
to $B-L$ and therefore $B-L=0$ does not imply $B=B+L=0$. This is contrary to
previous results and is important, since it was assumed that due to the
sphalerons one either requires early $B-L\not=0$ baryogenesis
or late baryogenesis.

(3){$\bf B-L=0$}
In models where $B-L$ is conserved at all temperatures one must
impose the additional constraint $B-L=0$. For $T\gsim T_C$, this
additional equation for the chemical potentials immediately gives
$B=L= B+L =0$, as well as $\mu_{uL}=\mu_-= \mu=0$. However, we
do not necessarily have $\mu_i=0 $.

If the initial conditions are such that $L_i=0$, and if the only
lepton-number violating interactions are
sphaleron processes, then the $ \mu_i$ will be equal and the
vanishing of $\mu$ implies the vanishing of $\mu_i$.
But if there is a pre-existing asymmetry in a
given $L_i$ channel, then $L_i- L_j$ conservation means that the vanishing of
$L$ can come about only through a cancellation between different,
non-zero, $\mu_i$. The non-zero $\mu_i$ imply a  non-zero
$\Delta\mu$ which will {\it regenerate} the lepton- and
baryon-asymmetry, similar to models discussed in
\cite{leptoruss}. This follows because, imposing $B-
L=0$, we can express $\mu_{uL}$ by $\Delta\mu$  and
\begin{eqnarray}
B &=& \left[
- \frac { \left(4N- 2\Delta_u +\frac {4N (2N-
\Delta_u)  } {2\alpha_W +b\alpha_-} \right) \left( {2{\alpha_W+b
\alpha_-}+ {3N}}\right)} {(13N-2 \Delta_u) ({\alpha_W+
\frac{b}{2}\alpha_-}) + {6N(2N-\Delta_u)}}
\right. \nonumber \\
&& + \frac{2N}{2\alpha_W+ b\alpha_-}
\left. \right] \Delta\mu,
\label{eq:barylow}
\end{eqnarray}
as well as $B=L=\half(B+L)$. In this case $\Delta\mu \not= 0$,
since $\mu_i \not=0$ and the baryon number  reappears {\it due}
to the sphaleron interactions converting $L_i$ number excess into
$B$ excess. We find the surprising result, that even when $B-L=0$ for
all temperatures we can have non-vanishing $B,L$ and $B+L$.

In the MSSM there are no further chemical potentials because; they are
all related to their SUSY partners'. This is because at the energies we
consider the neutralino (Majorana fermions!) have vanishing chemical
potential. For example the reaction ${\tilde e}^-\leftrightarrow e^-+
{\tilde\chi}^0$ in thermal equilibrium implies $\mu_{{\tilde e}}=\mu_e$.
Therefore the SM results, which depended on the number of independent
chemical potentials, remain qualitatively
the same. The main difference is that now three linear combinations of the
neutrino chemical potentials appear:
$$\mu,\Delta\mu,\quad \Delta{\tilde\mu}
= \mu-\sum{\tilde\alpha}_i\mu_i,$$
where
$${\tilde\alpha}_i=\alpha({\tilde m}_{ei}$$
denotes the mass corrections due to the sleptons now instead of the leptons
and we have assumed ${\tilde m}_{\nu i}={\tilde m}_{e i}$. Due to the larger
sfermion masses these can be
substantially larger. Taking $\Delta{\tilde\mu}\gg\Delta\mu$
we obtain in the case $B-L=0$
\beq
B= -24\frac{19+63{\tilde\alpha}+50{\tilde\alpha}^2}
{584+925{\tilde\alpha}+216{\tilde\alpha}^2} \Delta{\tilde\mu}.
\label{eq:new}
\eeq
And for ${\tilde\alpha}=1$, $B=-1.8\Delta{\tilde\mu}$. Finally we get
$L=B$ and $B+L=2B$.

\end{document}